\documentstyle[aps,preprint]{revtex}

\begin{document}

\title{Dielectron Cross Section Measurements in Nucleus-Nucleus Reactions
at 1.0 A$\cdot$GeV}

\author{(The DLS Collaboration)\\
R.J. Porter$^a$, S. Beedoe$^b$~\cite{sb}, R. Bossingham$^a$, M. Bougteb$^c$,
W.B. Christie$^d$~\cite{th}, J. Carroll$^b$,\\
W.G. Gong$^a$,
T. Hallman$^d$~\cite{th}, L. Heilbronn$^a$, H.Z. Huang$^{a,b}$, G. Igo$^b$,
P. Kirk$^e$,\\
G. Krebs$^a$, A. Letessier-Selvon$^a$~\cite{als},
L. Madansky$^d$, F. Manso$^c$, H.S. Matis$^a$, J. Miller$^a$,\\
C. Naudet$^a$~\cite{cjn}, M. Prunet$^c$, G. Roche$^{a,c}$, 
L.S. Schroeder$^a$, P. Seidl$^a$,\\  Z.F. Wang$^e$,
R.C. Welsh$^d$~\cite{rw}, W.K. Wilson$^{a,f}$,
A.Yegneswaran$^a$~\cite{ay}}

\address{$^a$ Lawrence Berkeley National Laboratory, University of
California, Berkeley, CA 94720, USA\\
$^b$ University of California at Los Angeles, CA 90095, USA\\
$^c$ Universit$\acute{e}$ Blaise Pascal/IN2P3, 63177 Aubi$\grave{e}$re
Cedex, France\\
$^d$ The Johns Hopkins University, Baltimore, MD 21218, USA\\
$^e$ Louisiana State University, Baton Rouge, LA 70803, USA\\
$^f$ Wayne State University, Detroit, MI 48201, USA}

\maketitle

\begin{abstract}
We present measured dielectron production cross sections for Ca+Ca, C+C,
He+Ca, and d+Ca reactions at 1.0 A$\cdot$GeV. Statistical uncertainties
and systematic effects are smaller than in previous DLS
nucleus-nucleus data. For pair mass $M\leq 0.35$ GeV/c$^{2}$: 1) the Ca+Ca 
cross section is larger than the previous DLS
measurement and current model results, 2) the mass spectra suggest 
large contributions from $\pi^{0}$ and $\eta$
Dalitz decays, and 3) $d\sigma/dM \propto$ A$_{P}\cdot$A$_{T}$. For $M > 
0.5 $ GeV/c$^{2}$ the 
Ca+Ca to C+C cross section ratio is
significantly larger than the ratio of A$_{P}\cdot$A$_{T}$ values. 
\end{abstract}



\newpage

Dielectrons produced in heavy-ion collisions are attractive probes for
studying dynamical properties of nucleus-nucleus interactions. The
$e^{+}e^{-}$ pairs do not undergo significant rescattering in the reaction,
thus the kinematics of the pairs retains information about their production.
This is of particular interest if the $e^{+}e^{-}$  pairs are 
produced by processes, such as pion 
annihilation,  that must occur 
during in the hot, dense phase of the collisions. Use of this probe has 
produced interesting results at both
Bevalac~\cite{Roche} and SPS~\cite{Agak} energies. We present in this letter 
the latest
measurements of dielectron production from the Dilepton Spectrometer 
(DLS)
Collaboration in nucleus-nucleus  reactions at a beam kinetic energy of 1.0
A$\cdot$GeV.

The DLS collaboration has previously reported on dielectron production in
several colliding systems~\cite{Roche,DLSO,HWH}. The first generation DLS 
data
from p+Be, Ca+Ca, and Nb+Nb~\cite{Roche,DLSO} reactions provided the 
first
observations of dielectrons produced at Bevalac energies. Early calculations
suggested that such data could be dominated by contributions from
$\pi^{+}\pi^{-}$ annihilation~\cite{Kap,Xia}. Subsequent models of AA
collisions in this energy regime~\cite{Wolf,Gudima,Cassing} calculated that
$e^{+}e^{-}$ pairs of invariant mass below about \mbox{0.4 GeV/c$^{2}$} are
produced primarily from conventional hadronic sources, such as $pn$ 
bremsstrahlung
and Dalitz decay processes ($\pi^{0}$, $\Delta$, and $\eta$), but that
contributions from $\pi^{+}\pi^{-}$ annihilation were needed to explain the
Ca+Ca data at higher pair masses. Models that focus on density induced 
changes
in the $\rho$--meson mass provide alternative descriptions of the pair yield
at the higher masses~\cite{Chan,Ko}. Within the limited statistics of the
first generation DLS data, it was not possible to distinguish among the models 
that
provided results for specific DLS measurements. 

After improvements to the DLS apparatus~\cite{Yegs,RJP,MP}, a second
generation of measurements was obtained: first from p+p and p+d reactions 
at a
number of energies~\cite{HWH}, and then from the Ca+Ca, C+C, He+Ca, and 
d+Ca
reactions presented in this letter. Each of these data sets contain
significantly more pairs than earlier DLS data. To increase our sensitivity to
the effects of multiple hadronic interactions (e.g. $\pi^{+}\pi^{-}$
annihilation and multi-step resonance excitation), the nucleus-nucleus
reactions were chosen to have different numbers of participant nucleons, but
identical isospin and similar internal nuclear motion. 

A description of the DLS apparatus has been published~\cite{Yegs}, and the
analysis procedures used here are discussed in Ref.~\cite{RJP}. The new 
procedures lead to a mass-independent mass resolution for these data of 
10\%, 
and smaller bin widths for M $\leq$ 0.2 GeV/c$^{2}$. No
explicit requirement on impact parameter (e.g. multiplicity) is imposed on
these data. The cross
sections presented here are evaluated in a three-dimensional, binned array of
invariant mass ($M$), transverse momentum ($p_{t}$), and laboratory 
rapidity
($y$). The data are available from the authors both as 3-D tables and as
projected spectra. A corresponding filter, which is necessary for comparisons 
with
theoretical results and different from filters used for other data 
sets, is also available. 
            
Early in the second generation of the DLS program, measurements showed a
previously unrecognized trigger inefficiency due to instantaneous rates much
larger than the well controlled average rates. The microscopic duty factor of
the beam, the instantaneous rates, and inefficiency varied on a time scale of
hours. The efficiency could be reliably monitored only by the yield of the
pairs themselves. A re-analysis of the first generation 4.9 GeV p+Be data
found a correction factor of $\sim$5 for the rate-dependent 
losses~\cite{Boug}. 
Because of the unstable nature of this problem, it is
unwarranted to assume that other first generation DLS data have the same
correction factor. Although they do show evidence for significant
rate-dependent inefficiencies, they lack sufficient information for
calculation of appropriate correction factors.  We suggest that the first
generation data no longer be used for comparison with theory. In the second
generation data, the rate-dependent losses were greatly reduced by
improvements to the electronics and the beam monitoring systems. For each
data set, we measured the rate dependence of the pair yield to permit
extrapolation to zero rate. After these improvements, a measurement of the
differential cross section for p+p elastic scattering at 1.27 GeV (made
concurrently with acquisition of p+p dielectron data) produced results in
agreement with the known cross section~\cite{Wilson2}, and gave rate-dependent
inefficiencies of $\leq$ 15\%. In the dielectron data reported here, run-by-run
corrections for such losses ranged between 10--45\%. 

Table I lists the data sets with the corresponding beam energies, pair
statistics and absolute normalization uncertainties. The uncertainties are
dominated by variations that result from using several methods to calculate
the corrections for the rate-dependent efficiency. Cross sections were
calculated from a single correction method, thus when comparing two data sets,
the appropriate relative uncertainty is $\sim$10--15\%. 

The mass dependence of the differential cross sections is shown in Fig.\ 1.
The systematic errors that are relevant to the shape of the spectra are
displayed in this figure as added linearly to the statistical errors. These
point-by-point systematics are independent of the normalization errors
given in Table I. They are obtained from studies of the combinatoric 
background and studies
of those acceptance corrections that were 
made in regions of phase space where the
acceptance changes rapidly.  Other effects, such as hadron contamination of
the $e^{+}e^{-}$ sample, are negligible compared to the systematics shown. For
masses below 0.2 GeV/c$^{2}$, an acceptance calculation with small bin widths
has changed the shape of the data. (Henceforward we refer always to results
within the DLS acceptance.) 

Fig.\ 1a displays the present data for the Ca+Ca cross section and the results
of several calculations. The dotted line is from a BUU model~\cite{Cassing}
and shows the general trend of models that adequately represent our previous
Ca+Ca data. For the present data the integrated cross section ($M >$ 0.2
GeV/c$^{2}$) is $\sim$7 times larger than both the earlier data and the model
results. We attribute the difference between our two measurements to the
uncorrected trigger inefficiency of the first generation DLS data. The
agreement between models and the present data remains reasonable above 0.6
GeV/c$^{2}$. In the mass range 0.2--0.4 GeV/c$^{2}$, however, the models 
predict pair yields that are dominated by $\Delta$ and $\eta$ Dalitz decays 
but are significantly lower than our measurement. The shape of
the yield is nonetheless quite similar to that of the $\eta$-component of the
BUU model shown in the dashed curve.

To emphasize the information contained in the spectrum shape, we have used a
simple model of $\pi^{0} $ and $\eta$ production to calculate the mass spectra
of the Dalitz decays. (Details of the model are discussed below.) The shapes
of these spectra are insensitive to the parameters of the model, and the shape
of the $\eta$ spectrum is in good agreement with that of the BUU model.
Fitting the Ca+Ca data (0.05 GeV/c$^{2}\leq M \leq$ 0.375 GeV/c$^{2}$) with
adjustable amplitudes of these two shapes yields the dashed curve shown in the
figure, with P($\chi^{2}_{\nu}$)$\simeq$5\%. The low probability of this fit
may be due to the contributions from other sources. The overall agreement
between curve and data shows that in this mass range the summed yield of pairs
from those sources is either similar in shape to that from the $\pi^{0}$ and
$\eta$ mesons, or relatively small, or slowly varying. 

As shown by the solid curves in the other panels of Fig.\ 1, the other data
sets also are well represented (for 0.05 GeV/c$^{2}\leq M \leq$ $\sim$0.4
GeV/c$^{2}$) by the same fitting procedure. We find that the fitted amplitude
for each component, as well as the ratio of the cross sections, scale as the
product of the projectile and target nucleon numbers, A$_{P}\cdot$A$_{T}$. For
the four reactions, the integrated cross sections (M $\leq$ 0.35 GeV/c$^{2}$)
scale as (A$_{P}\cdot$A$_{T}$)$^{\alpha}$ with $\alpha=1.06 \pm 0.01 \pm $0.02
(sys). 

A more direct comparison between the Ca+Ca and C+C data was obtained using the
ratio of the cross sections as a function of the pair mass. This ratio, shown
in Fig.\ 2, reveals two striking features. The first is that the ratio is
independent of pair-mass for $M\leq 0.4 $GeV/c$^{2}$. Fitting the ratio to
$d\sigma/dM\propto$ (A$_{P}\cdot$ A$_{T}$)$^{\alpha}$, gives $\alpha=1.01 \pm
0.03 \pm $ 0.04 (sys) - indicated by the line in the figure. This value is not
inconsistent with the calculations of Ref.~\cite{Wolf} for pairs from $\eta$
decay ($\alpha=0.87 \pm 0.1$) and $\Delta$ decay ($\alpha=0.95 \pm 0.1$)
produced in symmetric reactions ranging from Ca+Ca to Au+Au, and similar
behavior has also been found for sub-threshold K$^{+}$ production at 1.0
A$\cdot$GeV~\cite{elmer}. Such a large value of $\alpha$ is unexpected for
pairs from $\pi^{0}$ Dalitz decays, which are expected to dominate the low
mass region $M\leq 0.2 $GeV/c$^{2}$. Because of the shape of the DLS
acceptance, however, these low mass pairs are concentrated at rapidities
$\geq$ $y_{\rm beam}$, and this may produce their strong 
A$_{P}\cdot$A$_{T}$ dependence.

The second feature is the increase in the ratio for M $\ge$ 0.5 GeV/c$^{2}$,
where fitting gives $\alpha=1.40 \pm 0.13 \pm$ 0.04 (sys).  The data in Fig.\
1 suggest that this value of $\alpha$ is due to high-mass contributions to the
Ca+Ca data in addition to those producing the $\eta$-like shapes seen in the
other systems.  Although some models have suggested that these pairs are
primarily from $\pi^{+}\pi^{-}$ annihilations, Ref.~\cite{Wolf} finds that the
annihilation process should have $\alpha=1.00 \pm $0.05\, significantly
different from our observation. Calculations of density-dependent effects of
the medium on the pion propagator~\cite{Chan} show a structure at $\sim$0.5
GeV/c$^{2}$ in the $\rho$ mass spectrum which could be contributing these
additional pairs. Note that although these additional contributions could be
from a source that produces pairs only with M$ \ge$ 0.5 GeV/c$^{2}$, the
present data do not rule out a source which contributes over a larger mass
range but is visible only where other contributions (e.g. $\eta$ Dalitz decay)
are small. 

Direct comparisons of mass spectra for other combinations of data sets are not
made because at higher pair masses the rapidity distributions of the
asymmetric systems are different from each other and from that of the
symmetric systems. In the low-mass range these differences are not observed. 

Other data relevant to $\eta$ production in nucleus-nucleus collisions comes
from measurements by the TAPS Collaboration of inclusive differential cross
sections for $\pi^{0}$ and $\eta$ production in $^{40}$Ar+$^{\rm nat}$Ca (and
other) reactions at 1.0 A$\cdot$GeV~\cite{Schwalb,Berg}. The data span only a
small rapidity interval near $y_{\rm cm}$. Meson production data in this
limited rapidity range alone can not be used to calculate the Dalitz-decay
yields of these mesons in the DLS acceptance because of the incomplete
kinematic information. Thus there can be no model-independent statement
regarding the fraction of the DLS pair yield that each of these mesons
produces. The TAPS group calculated their total cross sections from a model
where mesons were emitted isotropically from a thermal source at $y_{\rm cm}$.
They obtained the temperatures of these sources by fitting their measured
$M_{t}$-spectra. Both the magnitude and $M_ {t}$ dependence of these data are
reproduced by the BUU model~\cite{Cassing}. 

We have used the TAPS mid-rapidity data, inferred temperatures, and thermal
model to calculate the contributions of their measurements within the DLS
acceptance. This is the simple model referred to above. (We use a value of the
$\eta$ cross section 30\% larger than the TAPS value because our data was
taken at a higher beam energy~\cite{Metag}.) The results are the histograms of
Fig.\ 1a. The contribution from $\pi^{0}$-Dalitz accounts for most of the
Ca+Ca cross section below 0.15 GeV/c$^{2}$. The $\eta$-Dalitz result, which is
consistent with the $\eta$ component of the BUU calculation~\cite{Cassing}, is
approximately 10\% of the DLS yield near 0.25 GeV/c$^{2}$. 

Significant (1 + a cos$^{2}$$\theta$) anisotropy is observed in both charged
pion data~\cite{KWolf,Harris}, and sub-threshold K$^{+}$
production~\cite{elmer}. Because the DLS acceptance is small at $y_{\rm cm}$
and peaks at $y_{\rm beam}$, this angular distribution for meson production
gives larger Dalitz decays yields than one which is isotropic. Our
calculations show that the $\eta$ yield increases more than that from the
$\pi^{0}$. Without direct information on the angular distribution of $\eta$
production we cannot quantify the size of this effect. Nevertheless, given the
significant uncertainties in both data sets~\cite{tape}, we find that the TAPS
data are not inconsistent with $\eta$ Dalitz decay contributing as much as
50\% of the DLS pair production in the mass range 0.2-0.4 GeV/c$^{2}$. 

In conclusion, we have presented dielectron measurements from Ca+Ca, C+C,
He+Ca, and d+Ca collisions at 1.0 A$\cdot$GeV.  We find that the Ca+Ca cross
section system is significantly larger than the original DLS measurement and
the model calculations based on it. The low-mass cross sections from the four
data sets reveal a mass-independent scaling of $d\sigma/dM \propto$
A$_{p}\cdot$ A$_{t}$ suggesting similar dynamics for the dominant source
mechanisms in reactions ranging from d+Ca to Ca+Ca.  The shape of the low-mass
spectra can be approximated with the estimated pair distributions from
$\pi^{0}$ and $\eta$ Dalitz decays. At higher pair mass, the ratio of Ca+Ca to
C+C cross sections is much larger than the A$_{p}\cdot$ A$_{t}$ ratio,
indicating that a density-dependent mechanism(s) may be exhibited in this mass
region.

The authors thank H. L$\ddot{o}$hner for his correspondence on the TAPS
measurement and acknowledge useful discussions with V. Koch and S Klein. We
also thank L. Bergstedt, L. Dean, D. Magestro, T. Meade, and L. Risk for their
help during the experiment.  We thank Al Smith for performing the beam ion
chamber calibrations and the Bevalac staff for their support. 

This work was supported by the Director, Office of Energy Research, Office of
High Energy and Nuclear Physics, Nuclear Physics Division of the U.S.
Department of Energy under contracts No. DE-AC03-76SF00098, No.
DE-FG03-88ER40424, No. DE-FG02-88ER40413

\begin{table}
\caption
{DLS dielectron data sets. "Comb" denotes the measured combinatoric
background, and "Sys" refers to the uncertainty in the absolute normalization.
Note the different beam kinetic energy for the $^{2}$H+$^{\rm nat}$Ca data.} 
\label{tab:one} 
\begin{tabular}{cccccc}
System  & E$_{beam}$ &  & Pair Yields   &      & \\
        & (A$\cdot$GeV)& $e^{+}e^{-}$  & Comb  & Net Pairs  & Sys  \\
\hline
$^{40}$Ca+$^{\rm nat}$Ca        & 1.04 & 12800  &  8102  &  4698 $\pm$ 145  &
$\pm$ 30\% \\
$^{12}$C+$^{\rm nat}$C & 1.04& 4760  &  1919  &  2841 $\pm$ 82  &
$\pm$30\% \\
$^{4}$He+$^{\rm nat}$Ca & 1.04 &  1929  &  487  &  1442 $\pm$ 49  &
$\pm$30\%
\\
$^{2}$H+$^{\rm nat}$Ca & 1.06 &  1828  &  308  &  1520 $\pm$ 43  &
$\pm$40\%
\\
\end{tabular}
\end{table}

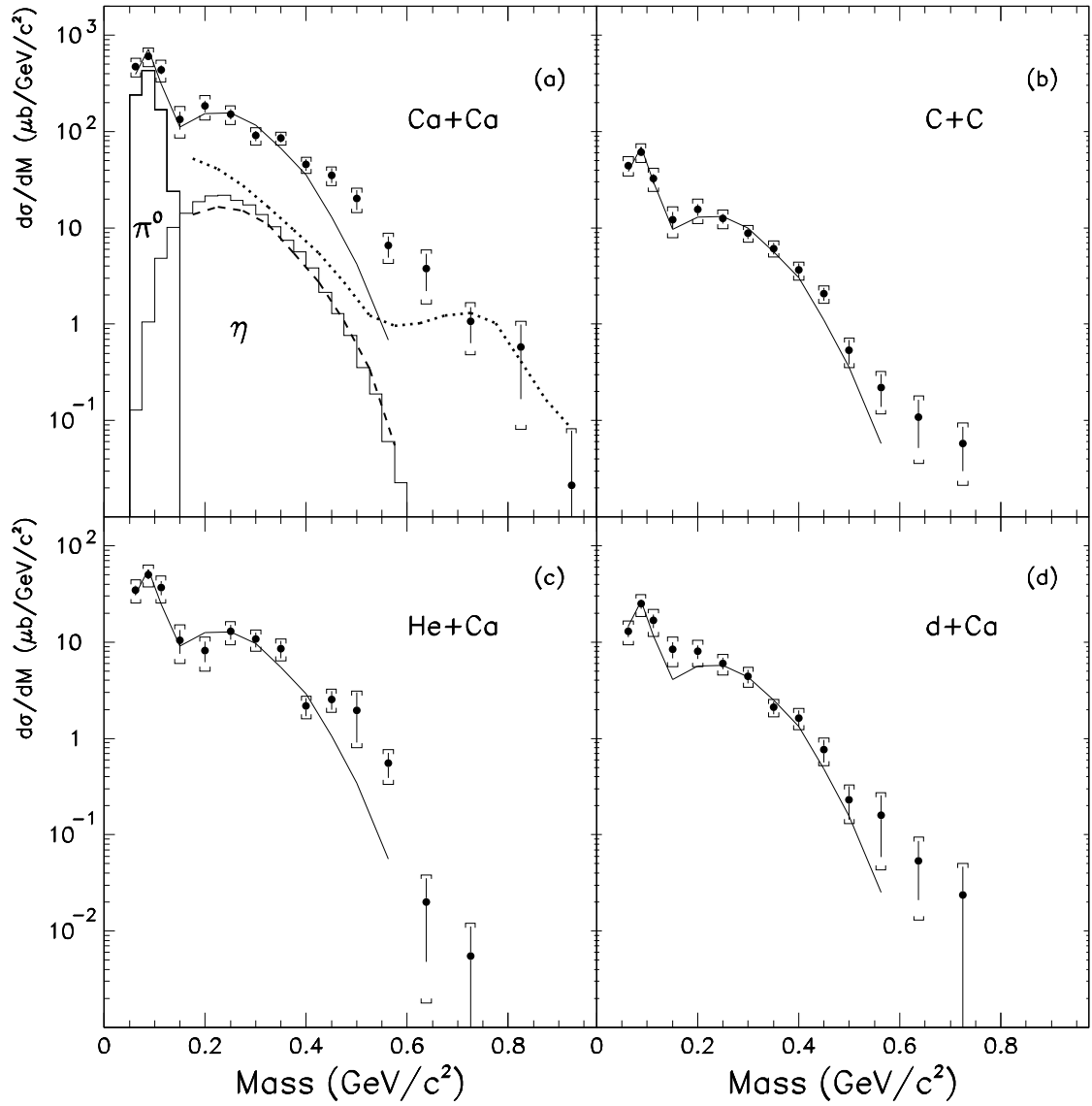
\begin{figure}
\caption{The DLS measurements of the dielectron cross sections from ($a$)
Ca+Ca, ($b$) C+C, ($c$) He+Ca, and ($d$) d+Ca reactions. Panel ($a$) also
contains the calculated signal from the BUU simulations of Ref. [9] (dotted 
line) and histograms showing the $\pi^{0}$ and $\eta$ decay 
contributions as estimated from the TAPS measurements and an isotropic thermal
model. The dashed line represents the $\eta$ component of the BUU calculation.
The solid lines in all four panels show our fit to the low-mass
data using the $\pi^{0}$ and $\eta$ decay estimates (histograms) with
adjustable normalizations.} 
\label{fig:one}
\end{figure}

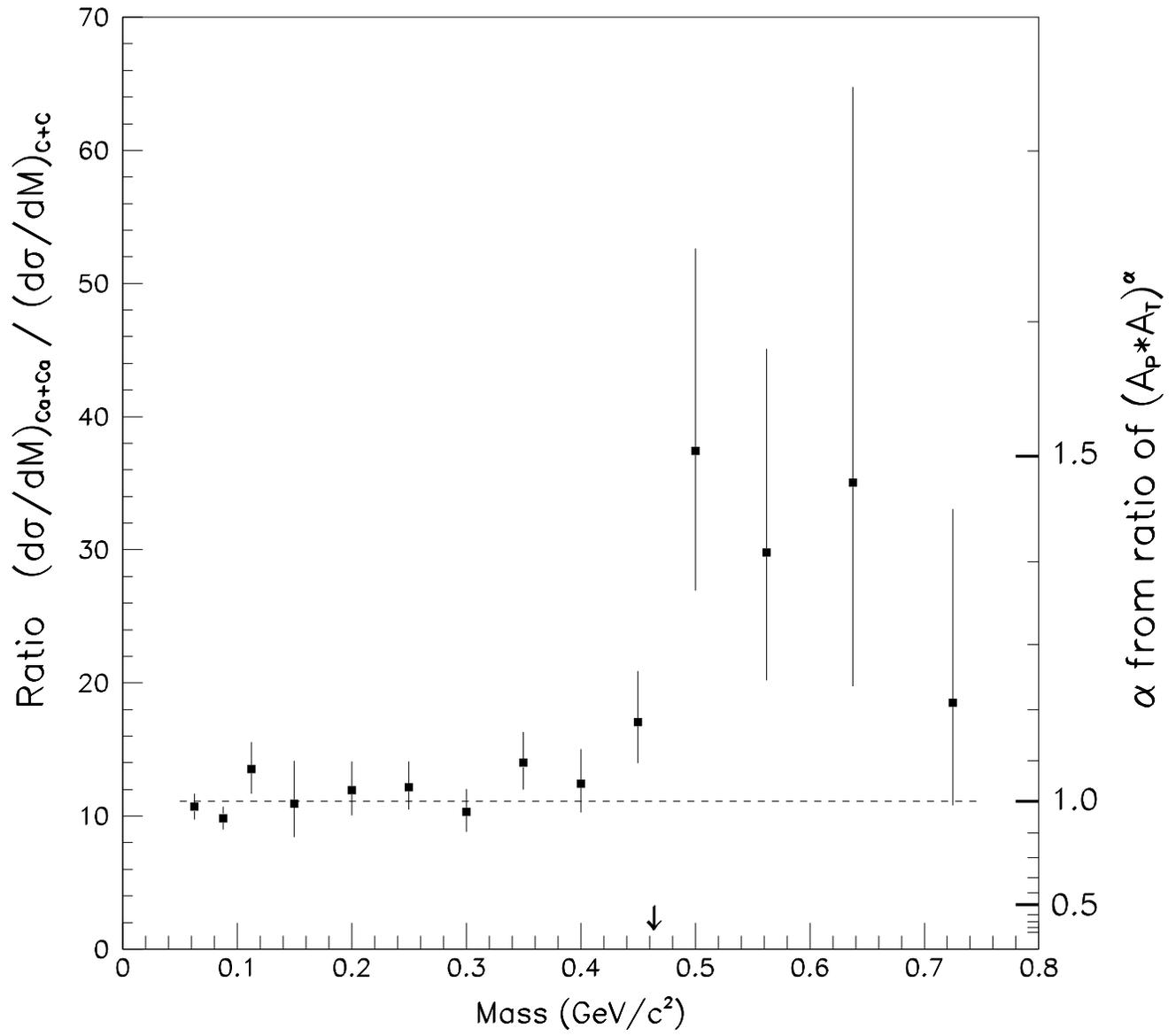
\begin{figure}
\caption{The ratio of the cross sections from Ca+Ca and C+C collisions as a
function of pair mass. The line on the plot is the fitted value of $\alpha$
from an assumed form $d\sigma/dM\propto$(A$_{p}\cdot$ A$_{t})^{\alpha}$. The
arrow indicates the kinematic limit for pair production in NN collisions.
Asymmetric errors in the ratio occur where there are large fractional errors
in the denominator; they are evaluated by Monte Carlo sampling.} 
\label{fig:two}
\end{figure}


\begin{references}
\bibitem{tape} The dominant error from DLS is the 30\% overall 
normalization. 
The dominant errors from TAPS are a 65\% statistical error and a 
systematic error on the combinatoric subtraction that is not discussed in 
their paper. Since the TAPS $\eta$ measurement contains a signal-to-noise 
ratio of about 0.1, even a 5\% bias in the background could lead to a 
correction of 50\%.
\end{references}
\end{document}